\documentclass[aps,prl,preprintnumbers,twocolumn,groupedaddress,nofootinbib]{revtex4}

\usepackage{latexsym}
\usepackage{amsmath}
\usepackage{amsfonts}
\usepackage{graphicx}

\newcommand{\bq}{\begin{eqnarray}}
\newcommand{\eq}{\end{eqnarray}}
\newcommand{\eps}{\varepsilon}

\newcommand{\arxivdate}{February 14, 2017}

\begin{document}

\preprint{MITP/17-009}
\title{\boldmath{Simplifying differential equations for multi-scale Feynman integrals beyond multiple polylogarithms}}

\author{Luise Adams, Ekta Chaubey and Stefan Weinzierl}
\affiliation{PRISMA Cluster of Excellence, Institut f{\"u}r Physik, Johannes Gutenberg-Universit\"at Mainz, D-55099 Mainz, Germany}

\date{\arxivdate}

\begin{abstract}
In this paper we exploit factorisation properties of Picard-Fuchs operators to decouple differential equations for
multi-scale Feynman integrals.
The algorithm reduces the differential equations to blocks of the size of the order of the irreducible factors of
the Picard-Fuchs operator.
As a side product, our method can be used to easily convert the differential equations for Feynman integrals which
evaluate to multiple polylogarithms to $\eps$-form.
\end{abstract}

\maketitle

\section{Introduction}
\label{sec:intro}

Precision physics with heavy particles, like the Higgs boson, the top quark or the $W$- and $Z$-bosons, 
plays an important part in the current physics programme at the LHC 
and will become even more important in the upcoming high luminosity runs.
Precision physics requires that higher-orders in perturbation theory are taken into account.
There is a class of mostly massless processes, where the virtual corrections can be expressed in terms of multiple polylogarithms.
Loop integrals can be tackled with differential equations \cite{Kotikov:1990kg,Kotikov:1991pm,Remiddi:1997ny,Gehrmann:1999as,Argeri:2007up,MullerStach:2012mp,Henn:2013pwa,Henn:2014qga}. 
We denote by $\eps$ the parameter of dimensional regularisation.
If the differential equations can be transformed to an $\eps$-form \cite{Henn:2013pwa,Henn:2014qga}, the solution in terms of multiple polylogarithms is straightforward.
However, starting at two-loops, there are integrals which cannot be expressed in terms of multiple polylogarithms.
The simplest example is given by the two-loop sunrise integral 
with internal masses \cite{Broadhurst:1993mw,Caffo:1998du,Laporta:2004rb,Kniehl:2005bc,MullerStach:2011ru,Adams:2013nia,Adams:2014vja,Adams:2015gva,Adams:2015ydq,Bloch:2013tra,Bloch:2014qca,Bloch:2016izu,Remiddi:2013joa}, 
where functions related to elliptic curves occur.
In the corresponding system of differential equations one faces the situation, that at order $\eps^0$ a system of two coupled
differential equations occur, which cannot be transformed away.

If we now look at realistic scattering processes at next-to-next-to-leading order (NNLO) with massive particles 
it is not unusual that within one topology
we have several master integrals, coupled together at order $\eps^0$ by the differential equations.
We denote by $N$ the number of master integrals within a given topology.
For example, for $2 \rightarrow 2$ processes at NNLO topologies with up to $5$ master integrals may occur.
It would be highly prohibitive, if we had to solve at order $\eps^0$ a coupled system of $N$ differential equations.
There are indications that topologies with three or more master integrals can be decoupled into blocks of size $2\times2$ at
worst \cite{Adams:2015gva,Remiddi:2016gno,Adams:2016xah,Bonciani:2016qxi,vonManteuffel:2017hms}.
This raises the question if there is a systematic method 
which transforms a system into an equivalent system, 
where at order $\eps^0$ the differential equations split into smaller blocks.
In this letter we will give an algorithm for this task.

The basic idea is as follows: We first reduce a multi-scale problem to a single-scale problem with scale $\lambda$.
In a second step we pick a master integral $I$ and determine at order $\eps^0$ and modulo sub-topologies 
the maximal number of independent derivatives $I$, $(d/d\lambda) I$, ..., $(d/d\lambda)^{r-1} I$.
This defines a Picard-Fuchs operator of order $r$.
For $r<N$ the system decouples into a system of $r$ master integrals and $(N-r)$ master integrals.
Let us look at the sector with $r$ master integrals.
In a third step we factorise the Picard-Fuchs operator.
This will decouple the system into blocks of the size
of the order of the irreducible factors of
the Picard-Fuchs operator.
In a fourth step we reconstruct the multi-variable transformation matrix from the single-variable one.

Although our primary interest are integrals involving elliptic sectors, it should be noted that our approach provides
as a side product an algorithm to convert a multi-scale system, which has a solution in terms
multiple polylogarithms to an $\eps$-form.
In this respect it complements 
other methods \cite{Gehrmann:2014bfa,Argeri:2014qva,Lee:2014ioa,Prausa:2017ltv,Gituliar:2017vzm,Meyer:2016slj,Tancredi:2015pta,Ablinger:2015tua}.

\section{The method}
\label{sec:method}

Let us consider a set of master integrals $I_1$, ..., $I_N$ depending on kinematic variables $x_1$, ..., $x_n$.
We denote the ordered set of master integrals by the vector $\vec{I}=(I_1,...,I_N)$.
If the master integrals depend only on a single kinematic variable $x_1$, we have a single-scale problem.
For two or more kinematic variables ($n \ge 2$) we have a multi-scale problem.
We consider the master integrals in $D=2m-2\eps$ space-time dimensions, with $m\in{\mathbb Z}$ 
and $\eps$ being the dimensional regularisation parameter.
Integration-by-parts identities \cite{Tkachov:1981wb,Chetyrkin:1981qh} allow us to derive a system of differential equations
of Fuchsian type
\bq
 d\vec{I} & = & A \vec{I},
\eq
where $A$ is a matrix-valued one-form
\bq
 A & = & 
 \sum\limits_{i=1}^n A_i dx_i.
\eq
The matrix-valued one-form $A$ satisfies the integrability condition
\bq
 dA - A \wedge A & = & 0.
\eq
We assume that $A$ has an $\eps$-expansion
\bq
 A & = &
 \sum\limits_{j \ge 0} \eps^j A^{(j)} 
 \;\; = \;\;
 \sum\limits_{i=1}^n \sum\limits_{j \ge 0} \eps^j A_i ^{(j)} dx_i.
\eq
The differential equations are usually solved order by order in $\eps$.
A crucial role for solving the system is played by the first term $A^{(0)}$.
The higher terms $A^{(j)}$ (with $j\ge1$) 
only give additional integrations over expressions of lower order.
We therefore seek transformations, which simplify $A^{(0)}$.
Under a change of basis
\bq
 \vec{J} & = & U \vec{I},
\eq
one obtains
\bq
 d \vec{J}
 & = &
 \tilde{A} \vec{J},
\eq
where the matrix $\tilde{A}$ is related to $A$ by
\bq
 \tilde{A} & = & U A U^{-1} - U d U^{-1}.
\eq
The master integrals can be expressed in terms of multiple polylogarithms
if there is a transformation $U$ such that $\tilde{A}^{(0)}=0$.

The matrix $A$ has a natural lower block triangular form, which derives from the top topology and its sub-topologies, obtained
by pinching of propagators.
In the following we consider the top topology and we work modulo sub-topologies.
The inclusion of sub-topologies leads only to integrations over already determined terms.
Let us assume that the top topology has $N$ master integrals.
We are in particular interested in the case where no transformation $U$ exists, such that $\tilde{A}^{(0)}=0$.
Although it might seem at first sight that we face in this situation at order $\eps^0$ a coupled system of $N$
differential equations it is very often the case that the system decouples into blocks of smaller size.
In this letter we give a systematic method to decouple the system.

We first reduce the multi-scale problem to a single-scale problem.
Let $\alpha=[\alpha_1:...:\alpha_n] \in {\mathbb C} {\mathbb P}^{n-1}$ be a point in projective space.
Without loss of generality we work in the chart $\alpha_n=1$.
Following \cite{Bonciani:2016qxi},
we consider a path $\gamma_\alpha : [0,1] \rightarrow {\mathbb C}^n$, indexed by $\alpha$ and parametrised by 
a variable $\lambda$.
Explicitly, we have
\bq
\label{def_path}
 x_i\left(\lambda\right) & = & \alpha_i \lambda,
 \;\;\;\;\;\; 1 \le i \le n.
\eq
We then view the master integrals as functions of $\lambda$.
In other words, we look at the variation of the master integrals in the direction specified by $\alpha$.
For the derivative with respect to $\lambda$ we have
\bq
\label{diff_eq_lambda}
 \frac{d}{d\lambda} \vec{I}
 & = &
 B \vec{I},
 \;\;\;\;\;\;
 B \; = \;
 \sum\limits_{i=1}^n \alpha_i A_i.
\eq
The matrix $B$ has again a Taylor expansion in $\eps$:
\bq
 B & = & B^{(0)} + \sum\limits_{j>0} \eps^j B^{(j)}.
\eq
Let $I$ be one of the master integrals $\{I_1,...,I_N\}$.
Eq.~(\ref{diff_eq_lambda}) allows us to express the $k$-th derivative of $I$ with respect to $\lambda$
as a linear combination of the original master integrals.
We recall that we work modulo sub-topologies. We may even work modulo $\eps$-corrections by using
$B^{(0)}$ instead of the full matrix $B$.
We then determine the largest number $r$, such that the matrix which expresses 
$I$, $(d/d\lambda)I$, ..., $(d/d\lambda)^{r-1}I$ in terms of the original set $\{I_1,...,I_N\}$ has full rank.
Obviously, we have $r \le N$.
In the case $r < N$ we complement the set $I, (d/d\lambda)I, ..., (d/d\lambda)^{r-1}I$ by $(N-r)$ elements
$I_{\sigma_{r+1}}, ..., I_{\sigma_N} \in \{I_1,...,I_N\}$ such that the transformation matrix has rank $N$.
The elements $I_{\sigma_{r+1}}, ..., I_{\sigma_N}$ must exist, since we assumed that the set $\{I_1,...,I_N\}$
forms a basis of master integrals for this topology.
The basis $\{I, (d/d\lambda)I, ..., (d/d\lambda)^{r-1}I, I_{\sigma_{r+1}}, ..., I_{\sigma_N} \}$ decouples the system 
into a block of size $r$, which is closed under differentiation at order $\eps^0$ modulo sub-topologies
and a remaining sector of size $(N-r)$.

Let us now investigate under which conditions the block of size $r$ can be decomposed further.
We recall that $r$ is the largest number such that $I, (d/d\lambda)I, ..., (d/d\lambda)^{r-1}I$
are independent.
It follows that $(d/d\lambda)^rI$ can be written as a linear combination of $I, (d/d\lambda)I, ..., (d/d\lambda)^{r-1}I$.
This defines the Picard-Fuchs operator $L_r$ for the master integral $I$ with respect to $\lambda$:
\bq
\label{def_picard_fuchs}
 L_{r} I & = & 0,
 \;\;\;\;\;\;
 L_r \; = \; \sum\limits_{k=1}^r R_k \frac{d^k}{d\lambda^k},
\eq
where the coefficients $R_k$ are rational functions in $\lambda$ and we use the normalisation $R_r=1$.
Note that the zero on the right-hand side of eq.~(\ref{def_picard_fuchs}) is understood modulo sub-topologies and modulo terms of order $\eps$.
Using always $B^{(0)}$ instead of $B$ ensures that $L_r$ is independent of $\eps$.
The Picard-Fuchs operator is easily obtained by a transformation to the basis $I, (d/d\lambda)I, ..., (d/d\lambda)^{r-1}I$.
In this basis the $r \times r$-matrix $\tilde{B}$ has the form
\bq
 \left( \begin{array}{ccccc}
 0 & 1 & ... & 0 & 0 \\
 && ... && \\
 0 & 0 & ... & 0 & 1 \\
 - R_0 & -R_1 & ... & -R_{r-2} & -R_{r-1} \\
 \end{array} \right).
\eq
It is very often the case that the operator $L_r$ factorises \cite{MullerStach:2012mp}:
\bq
\label{factorisation}
 L_r
 & = &
 L_{1,r_1} L_{2,r_2} ... L_{s,r_s}, 
\eq
where $L_{i,r_i}$ denotes a differential operator of order $r_i$.
Clearly, we have $r_1+...+r_s=r$.
The factorisation in eq.~(\ref{factorisation}) can be obtained with the help of standard computer algebra systems.
For example, Maple offers the command ``DFactor''.
The factorisation in eq.~(\ref{factorisation}) can be used to convert the system of differential equations
at order $\eps^0$ into a block triangular form
with blocks of size $r_1$, $r_2$, ..., $r_s$.
A basis for block $i$ is given by
\bq
 J_{i,j} & = &
 \frac{d^{j-1}}{d\lambda^{j-1}} L_{i+1,r_{i+1}} ... L_{s,r_s} I,
 \;\;\;\; 1 \le j \le r_i.
 \;\;\;
\eq
Let us denote $\vec{J}=(J_{1,1},...,J_{1,r_1},J_{2,1},...,J_{s,r_s})$.
Expressing the elements of $\vec{J}$ in terms of the original integrals $\vec{I}$ defines a transformation matrix
\bq
 \vec{J} & = & V \vec{I}.
\eq
$V$ is a function of the parameters $\alpha$ and of $\lambda$:
\bq
 V & = & V\left(\alpha_1,...,\alpha_{n-1},\lambda\right).
\eq
We recall that we work in the chart $\alpha_n=1$. 
Setting 
\bq
 U & = & V\left(\frac{x_1}{x_n},...,\frac{x_{n-1}}{x_n},x_n\right)
\eq
gives the transformation in terms of the original variables $x_1$, ..., $x_n$.
Let us mention that there might be terms in the original $A$, which map to zero in $B$ for the class of paths considered
in eq.~(\ref{def_path}).
These terms are derivatives of functions being constant on lines through the origin.
An example is given by
\bq 
 d \ln Z\left(x_1,...,x_n\right),
\eq
where $Z(x_1,...,x_n)$ is a rational function in $(x_1,...,x_n)$ and homogeneous of degree zero in $(x_1,...,x_n)$.
On the one hand these terms don't contribute if we integrate the differential equation along the paths of eq.~(\ref{def_path}).
On the other hand, these terms are in many cases easily removed by a subsequent transformation.

\section{The case of linear factors}
\label{sec:polylogs}

Let us consider the special case, where the Picard-Fuchs operator $L_r$ factorises completely into linear factors:
\bq
\label{factorisation_linear}
 L_r
 & = &
 L_{1,1} L_{2,1} ... L_{r,1}, 
\eq
with
\bq
 L_{i,1} & = & 
 \frac{d}{d\lambda} + R_{i,0}.
\eq
$R_{i,0}$ is a rational function in $\lambda$. Then it is possible to construct a transformation such that $\tilde{A}^{(0)}=0$.
We first set
\bq
 J_{i,1} & = &
 \exp\left( \int\limits^\lambda d\tilde{\lambda} R_{i,0} \right) L_{i+1,1} ... L_{r,1} I,
\eq
This transforms the system to a form 
\bq
 \frac{d}{d\lambda} \vec{J}
 & = & \tilde{B} \vec{J},
\eq 
where the $\eps^0$-term $\tilde{B}^{(0)}$ is lower triangular with zeros on the diagonal.
The possible non-zero entries in the lower triangle of $\tilde{B}^{(0)}$
are easily removed. It is sufficient to discuss the case of a $2 \times 2$-matrix:
For
\bq
 \tilde{B}^{(0)}
 & = &
 \left(\begin{array}{cc} 
  0 & 0 \\
  N & 0 \\
 \end{array} \right)
\eq
the transformation
\bq
 \tilde{V}^{(0)}
 & = &
 \left(\begin{array}{cc} 
  1 & 0 \\
  f & 1 \\
 \end{array} \right)
\eq
with
\bq
 f & = &
 - \int\limits^\lambda d\tilde{\lambda} N
\eq
removes the term in the lower left corner.
The function $N$ has a partial fraction decomposition in $\lambda$ and we use this technique to remove all terms
which are polynomials in $\lambda$ or poles of order $2$ and higher.
Single poles integrate to logarithms and indicate that our basis elements have non-uniform weight.
These are removed by rescaling the master integrals by $\eps$-dependent pre-factors, such that the master integrals
have uniform weight.
In summary, this gives an easy method to convert a system, 
where every Picard-Fuchs operator factorises into linear factors, into
$\eps$-form.

\section{Examples}
\label{sec:examples}

Let us now look at a few examples.
\begin{figure}
\begin{center}
\includegraphics[scale=1.0]{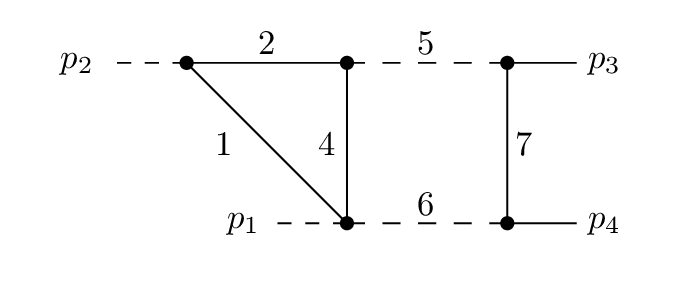}
\end{center}
\caption{
A two-loop four-point integral with six propagators.
}
\label{fig_example1}
\end{figure}
As a first example we consider a two-loop four-point integral with six propagators, shown in fig.~(\ref{fig_example1}).
Internal solid lines correspond to a mass $m$, dashed lines to mass zero.
The external momenta are on-shell, $p_1^2=p_2^2=0$ and $p_3^2=p_4^2=m^2$.
The Mandelstam variables are defined by $s=(p_1+p_2)^2$ and $t=(p_2+p_3)^2$.
The integral depends on two dimensionless variables $x_1$ and $x_2$, which we may choose as \cite{Fleischer:1998nb,Kotikov:2007vr,Bonciani:2010ms,Henn:2013woa}
\bq
 s = - m^2 \frac{\left(1-x_1\right)^2}{x_1},
 & &
 t = -m^2 x_2.
\eq
The internal propagators are labelled by numbers as shown in fig.~(\ref{fig_example1}). 
We denote by $I_{\nu_1 \nu_2 \nu_3 \nu_4 \nu_5 \nu_6 \nu_7}$ the scalar integral, where propagator $j$ occurs to the power $\nu_j$.
The seven indices refer to a double-box integral with seven propagators.
We use the convention that a scalar propagator is given by $1/(-q_j^2+m_j^2)$.
We use the program ``Reduze'' \cite{Studerus:2009ye,vonManteuffel:2012np} to obtain an initial set of master integrals together
with the corresponding system of differential equations. 
The integral in fig.~(\ref{fig_example1}) has an elliptic sub-topology, obtained by pinching propagators $2$, $5$ and $6$.
The topology in fig.~(\ref{fig_example1}) has two master integrals, which we may take as
\bq
 \vec{I} & = & \left( \left(1+2\eps\right) I_{1101111}, I_{1101211} \right).
\eq
The pre-factor $(1+2\eps)$ in front of the first master integrals ensures that only $A^{(0)}$ and $A^{(1)}$ appear in the
$\eps$-expansion of $A$.
The Picard-Fuchs operator for $I=(1+2\eps) I_{1101111}$ is of order $2$ and factorises into linear factors:
\bq
\lefteqn{
 L_2 = } & & \nonumber \\
 & &
 \left(
  \frac{d}{d\lambda} 
  + \frac{1}{\lambda+1} + \frac{2\alpha_1}{\alpha_1 \lambda -1} 
  + \frac{2 \alpha_1 \lambda}{\alpha_1 \lambda^2 +1}
  - \frac{2 \alpha_1 \lambda}{\alpha_1 \lambda^2 -1}
 \right)
 \nonumber \\
 & &
 \times
 \left(
  \frac{d}{d\lambda} 
  - \frac{1}{\lambda} + \frac{1}{\lambda+1} + \frac{2\alpha_1}{\alpha_1 \lambda -1}
 \right).
\eq
We may therefore transform to a basis, where $\tilde{B}^{(0)}$ is lower triangular with zeros on the diagonal.
The entry on the lower left corner of $\tilde{B}^{(0)}$ has only single poles and is removed by rescaling the first master
integral with $(1+2\eps)$ and the second master integral by $\eps$.
This converts the system with respect to the variable $\lambda$ to $\eps$-form.
Going back to the original variables we find
\bq
 \tilde{A}
 & = &
 \left(\begin{array}{cc} 
  1 & 0 \\
  0 & 1 \\
 \end{array} \right)
 d \ln\left(\frac{x_1}{x_2}\right)
 +
 \eps \tilde{A}^{(1)}.
\eq
The $\eps^0$-term is easily removed by multiplying both master integrals by $x_2/x_1$.
In summary we find that with
\bq
 & &
 U =  
 \left( \begin{array}{cc}
 U_{11}
 &
 - \frac{\left(1+2\eps\right)\left(x_1-1\right)^3\left(x_2+1\right)^2}{2 x_1 \left(x_1+1\right)} \\
 \frac{\eps\left(x_2+1\right)\left(x_1-1\right)^2}{x_1} & 0 \\
 \end{array} \right), \nonumber \\
 && U_{11} = 
{\scriptstyle
 \frac{\left(1+2\eps\right) \left(x_1-1\right)\left(x_2^2 x_1 + x_2 x_1^2 + x_2 - x_1^2 + 3 x_1 -1 \right)}{2 x_1 \left(x_1+1\right)}
} 
\eq
the transformed system is given by
\bq
 \tilde{A}
 & = & 
 \eps \left[
  \left( \begin{array}{rr}
    2 & 0 \\
    0 & 0 \\
  \end{array} \right) d\ln\left(x_1+1\right)
 -
  \left( \begin{array}{rr}
    2 & 0 \\
    0 & 2 \\
  \end{array} \right) d\ln\left(x_1-1\right)
 \right. \nonumber \\
 & & \left.
 -
  \left( \begin{array}{rr}
    0 & 0 \\
    0 & 2 \\
  \end{array} \right) d\ln\left(x_2+1\right)
 +
  \left( \begin{array}{rr}
    0 & 0 \\
    -1 & 1 \\
  \end{array} \right) d\ln\left(x_1+x_2\right)
 \right. \nonumber \\
 & & \left.
 +
  \left( \begin{array}{rr}
    0 & 0 \\
    1 & 1 \\
  \end{array} \right) d\ln\left(x_1 x_2+1\right)
 \right].
\eq
We see that this topology can be transformed to $\eps$-form and does not introduce new elliptic integrations.

Let us now look at a more involved example.
\begin{figure}
\begin{center}
\includegraphics[scale=1.0]{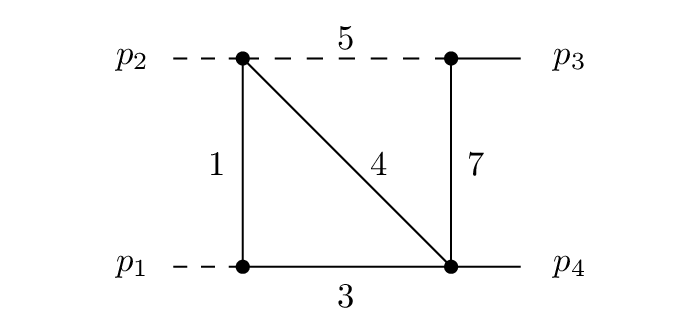}
\end{center}
\caption{
A two-loop four-point integral with five propagators.
}
\label{fig_example2}
\end{figure}
We consider the two-loop four-point integral with five propagators shown in fig.~(\ref{fig_example2}).
The kinematics is as in our first example.
This topology has five master integrals.
As our initial basis we take
\bq
 \vec{I} =  
 \left( \eps I_{1011101}, I_{2011101}, I_{1021101}, I_{1012101}, I_{1011201} \right).
\eq
Multiplying the first master integral by $\eps$ 
ensures that only $A^{(0)}$ and $A^{(1)}$ appear in the
$\eps$-expansion of $A$.
For the specific system under consideration this also decouples the first master integral $I_1=\eps I_{1011101}$
at order $\eps^0$ from the remaining ones.
We therefore have to consider only a $4 \times 4$-system.
Let us pick $I_2=I_{2011101}$. 
Working modulo $\eps$-terms, we find that already the third derivative of $I_2$ can be expressed as a linear combination of the lower ones.
Adding $I_5=I_{1011201}$ to $I_2, (d/d\lambda)I_2, (d/d\lambda)^2I_2$ will give a transformation matrix of full rank.
This decouples $I_5$ from the $3 \times 3$-system formed by $I_2, (d/d\lambda)I_2, (d/d\lambda)^2I_2$.
The Picard-Fuchs operator for $I_2$ is therefore of order $3$. It factorises into a second-order operator and a first-order operator:
\bq
 L_3 & = & L_2 L_1.
\eq
This decouples the $3 \times 3$-system into a $2 \times 2$-system and a $1 \times 1$-system.
The $2 \times 2$-system is irreducible.
When lifting the result from the single-scale case to the multi-scale case with the variables $\{x_1,x_2\}$ we again perform
an additional transformation, which removes $d\ln(x_1/x_2)$-terms.
In summary we are able to decompose the five master integrals for this topology at order $\eps^0$ in blocks of size
\bq
 1, 2, 1, 1.
\eq
The explicit expressions are longer, however we may display the structure of $\tilde{A}$. We have
\bq
 \tilde{A}
 & = & 
 \left( \begin{array}{ccccc}
 0 & 0 & 0 & 0 & 0 \\
 0 & * & * & 0 & 0 \\
 0 & * & * & 0 & 0 \\
 0 & * & * & 0 & 0 \\
 0 & * & * & 0 & 0 \\
 \end{array} \right)
 + \eps \tilde{A}^{(1)},
\eq
where $*$ indicates a non-zero entry.
In this example we see that $A^{(0)}$ cannot be transformed to zero. 
We find an irreducible $2\times2$-system at order $\eps^0$.
However, we achieved to simplify the original $5\times5$-system to smaller blocks.

In addition we have applied our method successfully to all sectors of the seven-propagator double-box integral, including
the top sector with seven propagators. This sector has five master integrals and decouples into blocks of size $1$, $2$, $1$ and $1$.

\section{Conclusions}
\label{sec:conclusions}

In this letter we presented an algorithm to simplify differential equations for multi-scale Feynman integrals.
We first reduced the problem to a single-scale problem and exploited then factorisation properties of the Picard-Fuchs operator.
This allows us to decouple the system at order $\eps^0$ into blocks of the sizes of the irreducible factors of the 
Picard-Fuchs operator.
We expect this technique to be useful for precision calculations.
A particular special case is given when all Picard-Fuchs operators factorise into linear factors.
In this case our method provides an easy algorithm to convert a multi-scale differential system into $\eps$-form.

\subsection*{Acknowledgements}

L.A. and E.C. are grateful for financial support from the research training group GRK 1581.

\begin{appendix}

\end{appendix}

\bibliography{/home/stefanw/notes/biblio}
\bibliographystyle{/home/stefanw/latex-style/h-physrev5}

\end{document}